\documentclass[usenatbib]{mnras}

\usepackage{graphicx}
\usepackage[space]{grffile}
\usepackage{latexsym}
\usepackage{amsfonts,amsmath,amssymb}
\usepackage{siunitx}
\usepackage{url}
\usepackage[utf8]{inputenc}
\usepackage{hyperref}
\hypersetup{colorlinks=false,pdfborder={0 0 0}}
\usepackage{textcomp}
\usepackage{longtable}
\usepackage{multirow,booktabs}
\usepackage{upgreek}
\usepackage{threeparttable}
\usepackage{caption}
\usepackage{color}

\DeclareMathSymbol{\ast}{\mathbin}{symbols}{"03}

\title[Microstructure revealed in FRB170827]{FRB microstructure revealed
  by the real-time detection of FRB170827}

\author[Wael Farah et al.]{\parbox{\textwidth}{W. Farah$^{1}$, C. Flynn$^{1,2}$,
    M. Bailes$^{1}$, A. Jameson$^{1,2}$, K. W. Bannister$^{3}$, 
    E. D. Barr$^{4}$, T. Bateman$^{5}$, S. Bhandari$^{1,2}$, M. Caleb$^{2,6,7}$, 
    D. Campbell-Wilson$^{5}$, S.-W. Chang$^{2,6}$, A. Deller$^{1,2}$, 
    A. J. Green$^{5}$, R. Hunstead$^{5}$, F. Jankowski$^{1,2,7}$,
    E. Keane$^{2,8}$, J.-P. Macquart$^{3,9}$, A. M\"oller$^{2,6}$, 
    C. A. Onken$^{2,6}$, S. Os{\l}owski$^{1}$, 
    A. Parthasarathy$^{1,2}$, K. Plant$^{1,10}$, V. Ravi$^{10}$, R. M. Shannon$^{1,11}$, 
    B. E. Tucker$^{2,6}$,
    V. Venkatraman Krishnan$^{1,2}$, C. Wolf\,$^{2,6}$} \\ \\ \\
\parbox{\textwidth}{$^1$Centre for Astrophysics and Supercomputing,
  Swinburne University of Technology, Mail H30, PO Box 218, VIC 3122,
  Australia\\
$^{2}$ARC Centre of Excellence for All-sky Astrophysics (CAASTRO)\\
$^3$ATNF, CSIRO Astronomy and Space Science, PO Box 76, Epping, NSW 1710, Australia\\
$^4$Max-Plank-Institute f\"{u}r Radioastronomie, Auf dem H\"{u}gel 69, D-53121 Bonn, Germany\\
$^5$Sydney Institute for Astronomy, School of Physics A28, University
  of Sydney, NSW 2006, Australia\\
$^6$Research School of Astronomy and Astrophysics, Australian National University, Canberra, ACT 2611, Australia\\
$^{7}$Jodrell Bank Centre for Astrophysics, School of Physics and Astronomy, The University of Manchester, Manchester M13 9PL, UK\\
$^{8}$SKA Organization, Jodrell Bank Observatory, Cheshire SK11 9DL,
  UK\\
$^{9}$International Centre for Radio Astronomy Research, Curtin University, Bentley, WA 6102, Australia\\
$^{10}$Cahill Centre for Astronomy and Astrophysics, MC 249-17, California Institute of Technology, Pasadena, CA 91125, USA\\
$^{11}$The Australian Research Council Centre of Excellence for Gravitational Wave Discovery (OzGrav)}
}

\date{Accepted 2018 April 21. Received 2018 April 20; in original form 2018 March 07 }

\pubyear{2018}

\begin{document}
\label{firstpage}
\pagerange{\pageref{firstpage}--\pageref{lastpage}}
\maketitle

\begin{abstract}
We report a new Fast Radio Burst (FRB) discovered in 
real-time as part of the UTMOST project at the Molonglo Observatory Synthesis Radio Telescope (MOST). 
FRB170827 is the first detected with our low-latency ($< 24$ s),
machine-learning-based FRB detection system. The FRB discovery was accompanied by the capture of 
voltage data at the native time and frequency resolution of the observing system, 
enabling coherent dedispersion and detailed off-line analysis, which have unveiled fine temporal and 
frequency structure. 
The dispersion measure (DM) of 176.80 $\pm$ 0.04 pc\,cm$^{-3}$, is 
the lowest of the FRB population. The Milky Way contribution along the line of sight is $\sim$ 40 pc\,cm$^{-3}$, leaving an excess DM of $\sim$ 145 pc\,cm$^{-3}$.
The FRB has a fluence $>$ 20 $\pm$ 7 Jy\,ms,  
and is narrow, with a width of $\sim$ 400 $\upmu$s at 10$\%$ of its maximum amplitude.
However, the burst shows three temporal components, the narrowest of which is $\sim$ 30 $\upmu$s, and a scattering timescale of $4.1 \pm 2.7$ $\upmu$s. The FRB shows spectral modulations on frequency scales of 1.5 MHz and 0.1 MHz. Both are prominent in the dynamic spectrum, which shows a very bright region of emission between 841 and 843 MHz, and weaker, patchy emission across the entire band. We show the fine spectral structure could arise in the FRB host galaxy, or its immediate vicinity. 


\end{abstract}


\section{Introduction} 
\indent Fast Radio Bursts (FRBs) form a class of extragalactic radio transients, with
approximately 30 published since \cite{Lorimer2007} reported the first.
The dispersion measures (DMs) of known FRBs currently spans the 
range $175 - 2600$ pc cm$^{-3}$ (FRBcat\footnote{\href{http://frbcat.org}
{\url{http://frbcat.org}}; visited 07/03/2018}; \citealt{FRBCat}), 
vastly exceeding the contribution of the Milky Way along their line of sight. 
FRBs have been detected at the Green Bank Telescope (GBT), the Parkes radio telescope, the Arecibo Observatory, the upgraded Molonglo Observatory Synthesis Telescope
(UTMOST) and the Australian Square Kilometer Array Pathfinder (ASKAP) 
(\citealt{Lorimer2007}; \citealt{Keane2012}; 
\citealt{Burke-Spolaor2014}; \citealt{Spitler2014}; \citealt{Petroff_realtime}; \citealt{Ravi2015}; \citealt{Champion2015}; 
\citealt{Masui2015}; \citealt{Kean_nature}; \citealt{Ravi2016_science}; \citealt{Petroff2017};
\citealt{Caleb_utmost}; \citealt{Bannister_etal}; \citealt{Bhandari2018}).

\cite{Bhandari2018} have recently estimated the FRB event rate from 
FRBs found in the HTRU \citep{HTRU} and SUPERB \citep{SUPERB} surveys at Parkes as 
$1.7^{+1.5}_{-0.9}\times10^3$\,FRBs\,($4\pi$ sr)$^{-1}$day$^{-1}$ above
$\sim2$ Jy\,ms. At 19 FRBs, this is the largest sample of FRBs
found with a single instrument. The authors show that there is no strong evidence that FRBs
are other than isotropically distributed on the sky, although the 
sample size remains small. \cite{Macquart2017} found 
that the slope of the cumulative source count distribution of Parkes FRBs is 
$\alpha=-2.6^{+0.7}_{-1.3}$, implying 
a non-uniform space density with distance or source population evolution. Larger 
samples are required to address this issue conclusively, and are expected to become available as 
facilities like ASKAP \citep{Bannister_etal} and CHIME \citep{CHIME} reach full FRB search capacity.    

Only one of the FRBs reported to date has been found to repeat 
(FRB121102; \citealt{Spitler2016}), with the emission showing no sign of an underlying periodicity. 
The opportunity for repeated targeting of this source has led to a precise localisation 
using radio interferometers, with FRB121102 pinpointed to a low-metallicity 
dwarf galaxy at a redshift of $z = 0.193$ \citep{Chatterjee2017, Tendulkar2017,Bassa2017}, 
co-located to within a region $\lesssim 12$ mas 
with a persistent radio source \citep{Marcote2017}. 
The properties of the host galaxy of FRB121102 are similar to those 
of the hosts of 
hydrogen-poor superluminous supernovae (SLSNe-I), leading \cite{Metzger2017} to 
propose that the repeated bursts from FRB121102 originate from a young magnetar 
remnant embedded within a young hydrogen-poor supernova remnant which would be at most 
a few decades old. \cite{Michilli2018} have recently shown that the bursts of FRB121102 
exhibit extreme Faraday rotation measures, implying that the source 
resides in a highly magnetised region.

Many theories have been formulated to explain FRBs, and can be 
broadly classified as non-cataclysmic and cataclysmic, depending on the fate of the 
progenitor. Non-cataclysmic theories include giant flares from magnetars 
\citep{Pen&Connor2015}, compact objects in young supernovae \citep{Metzger2017} and
supergiant pulses from extragalactic neutron stars \citep{Cordes&wasserman2015}. 
Cataclysmic events include scenarios such as neutron star mergers \citep{Totani2013} and 
``blitzars'' which 
occur when a spinning-down neutron star collapses into a black hole \citep{Falcke2014},
releasing the neutron star's magnetosphere. Cataclysmic models for the events are 
challenged by the existence of the repeating FRB, 
although it seems unlikely that FRB121102 is representative of the entire FRB population \citep{Palaniswamy2018}.

A key element to a better understanding of FRBs is to localise them to their host galaxies. 
We are currently limited to $\approx 10$ arcmin radius localisations of FRBs 
at Parkes, GBT and ASKAP, 
or narrow fan-beam localisations with UTMOST 
which are 5 arcsec $\times$ 1.2 degrees. 
Plans are currently afoot at UTMOST (the UTMOST-2D
project), ASKAP \citep{Bannister_etal}, MeerKAT, and the Very Large Array \citep{Law2015}, 
to achieve localisations ranging from a few square arcminutes 
to a square arcsecond or less.
The North-South arm of the Molonglo telescope, which crosses at right angles
relative to the East-West, is currently being fitted out to approximately
the same sensitivity (fluence $\approx 5$ Jy\,ms for a 1 ms burst) as the
East-West arm. 
The localisation that these facilities plan to deliver
in the coming 24 months is in the range of $\approx$ 10 arcsec down
to sub-arcsecond scales, which permits host galaxy identifications for FRBs depending on the source redshift \citep{Eftekhari2017}.

Contingent to source localisation at UTMOST is the capacity to do
real-time detection of FRBs. Due to the RFI (Radio Frequency Interference) environment at the site, 
it is advantageous to use a machine-learning-based system to avoid false 
triggers so that voltage data can confidently be recorded when FRB events occur. 
We have developed such a system, and the FRB reported here is our
first detection since the system sensitivity was improved substantially 
by hardware changes implemented in July 2017. Recording voltages preserves amplitude and phase information 
about the incident electromagnetic radiation. Hence, voltage capture not only leads to superior 
localisation of the FRB, but it also gives remarkable temporal and
frequency resolution relative to our now superseded system.  

In this paper we report the first FRB for which a voltage capture has been achieved after 
a real-time discovery. In section~\ref{Observations}, we describe Molonglo's observing 
setup, and the real-time pipeline. We report the discovery of FRB170827 in section~\ref{Discovery}. 
The analysis of the dynamic spectrum and pulse profile of the FRB is presented in section~\ref{analysis}. 
In section~\ref{Multiwavelength}, we present the conducted multiwavelength followup.
We discuss our results in section~\ref{discussion}.
\begin{figure}
  \begin{center}
    \includegraphics[angle=-90,scale=0.34]{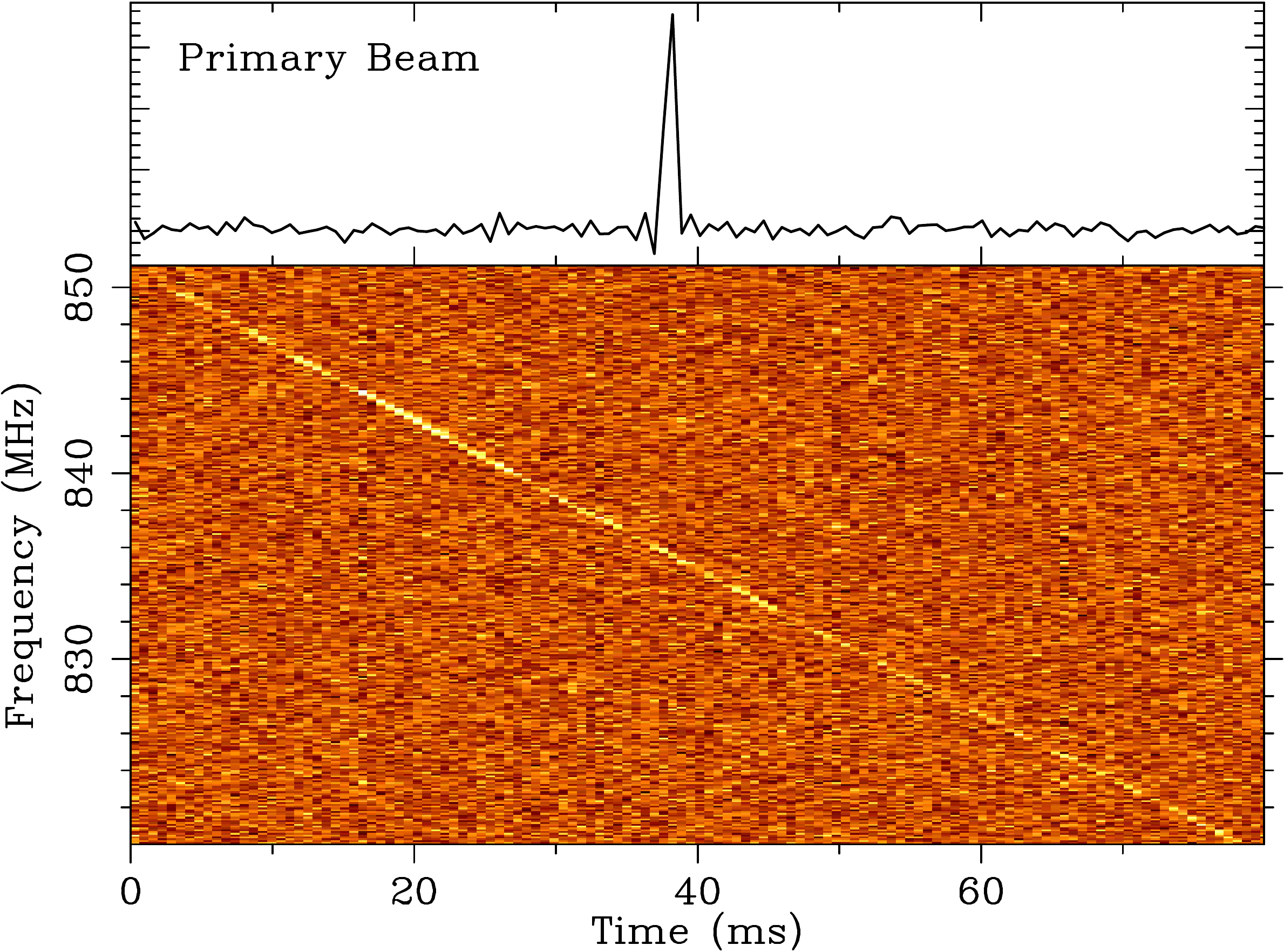}
    
    \vspace{0.2cm}
    
    \includegraphics[angle=-90,scale=0.33]{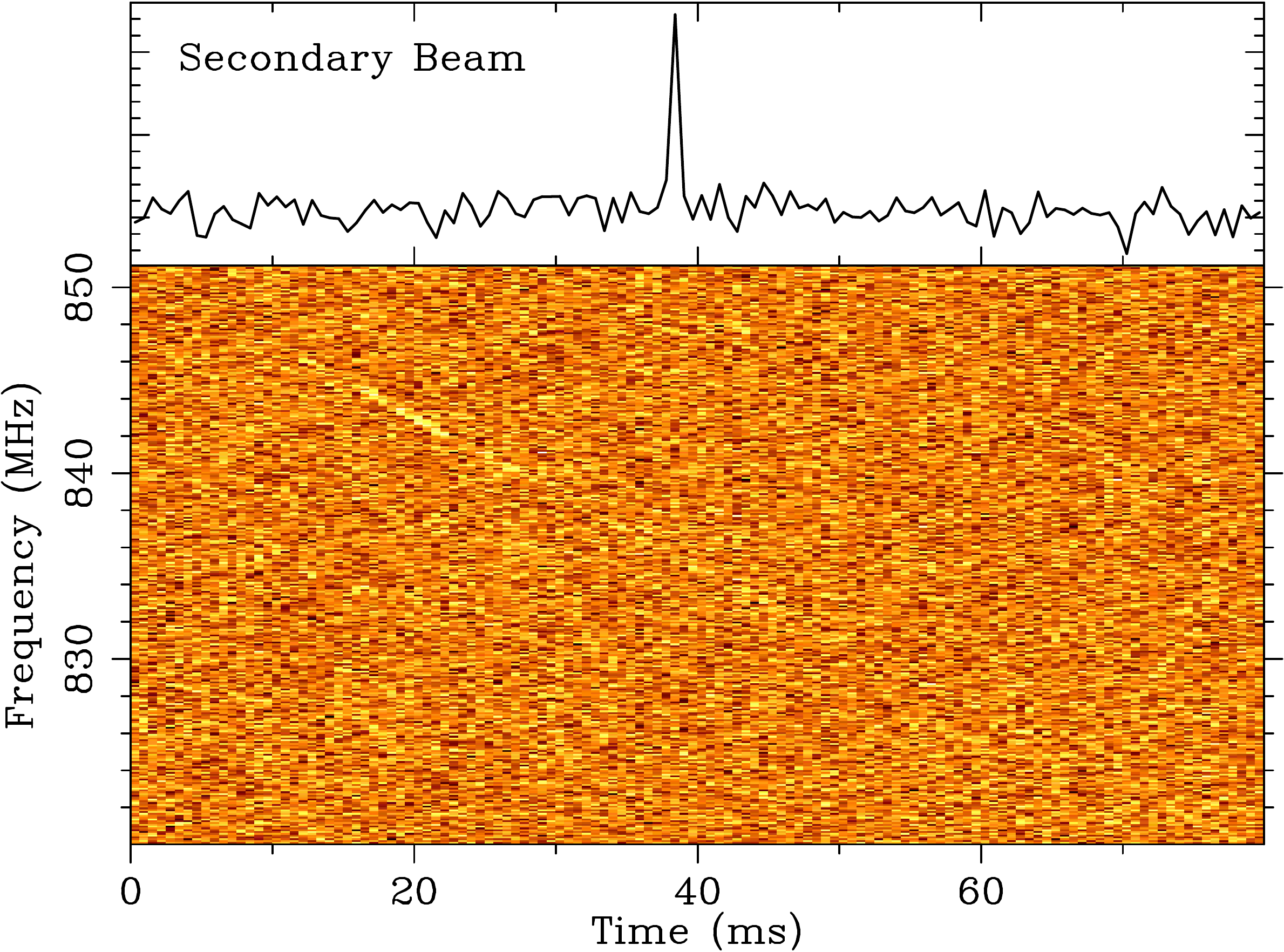}
        
    \caption{Top: Fan-beam of primary detection of FRB170827 
      at a resolution of 327.68\,$\upmu$s and 97.66\,kHz. 
      Bottom: adjacent fan-beam detection. The detection 
      of FRB170827 in only two fan-beams is consistent with a source originating 
      in the far field.}
    \label{frbdetection}
  \end{center}
\end{figure}

\section{Observations and the UTMOST data capture system}
\label{Observations}
\indent The UTMOST is an on-going project to transform MOST, 
a radio interferometer located
near Canberra, Australia, into an FRB finder \citep{utmost}. 
The 1.6 km extent of UTMOST reflector is divided into 352 ``modules'', where each module consists of 22 
ring antennas that select a single right circular polarisation and allow the module beam to be steered. 
The central observing frequency of Molonglo is 835 MHz and the total bandwidth is 31.25 MHz.
Since 2017 July, we have been operating the UTMOST telescope as a
transit facility, and steering of the array away from the meridian has
been sacrificed for better sensitivity.  
The mechanical system
responsible for phasing these antennas away from the meridian was
approaching end-of-life and had become unreliable when slewing, such
that the sensitivity had fallen well below that in early 2016 when
three FRBs were discovered \citep{Caleb_utmost}. With the switch to
transit mode, the telescope sensitivity improved approximately three-fold. 
The system is proving to be very
stable, with phasing calibration of the system required only a few
times weekly.

\subsection{Real-time analysis and voltage capture}

\indent The real-time FRB analysis system, at its core, comprises 
a GPU-based single-pulse processing software, \textsc{HEIMDALL}\footnote{\href{https://sourceforge.net/projects/heimdall-astro/}
{\url{https://sourceforge.net/projects/heimdall-astro/}}}. 
The input to \textsc{HEIMDALL} is the high resolution filterbank data (327.68 $\upmu$s, 97.66 kHz, 8 bit) for each of the 352 fan-beams: equally spaced tied array beams that cover the field of view of an individual UTMOST module.
\textsc{HEIMDALL} 
searches the fan-beam data streams, spread across 8 GPUs. 
The software performs dedispersion (from 0 to 2000 pc\,cm$^{-3}$)
and matched filtering (from 327.68 $\upmu$s to 83.886 ms), and reports time-stamped FRB candidates. 
Since part of the UTMOST operating spectrum is allocated to mobile phone communications, 
the vast majority of FRB candidates are artifacts, numbering tens of thousands per night. The system was 
thus extended by a supervised machine-learning algorithm, based on random forest \citep{Breiman2001}, 
in order to perform real-time candidate classification. The model was trained on single pulses from 
various pulsars, and on RFI-generated candidates, achieving an accuracy of 98.8\% (10-fold cross-validation).
To perform classification, the model is specified by predefined features, extracted from the 
input data stream of a given candidate. These features are DM and S/N agnostic, and  sufficient  
to characterise the RFI activity during an event, the noise statistics, and the validity of the candidate. 
The details of the pipeline will be presented in a subsequent paper. 

Application of the pipeline reduced the number of candidates to a manageable number, $\sim$10 a day, the first step to making voltage 
capture a possibility. Channelised voltages are buffered in Random Access Memory 
for 24 seconds, sufficient to perform real-time beam-forming, 
single pulse searching and classification. Upon a trigger, 
a voltage capture around an interesting event is performed, while accounting for dispersive delay and 
allowing a narrow buffer window for baseline estimation. The captured data are the channelised, critically sampled 
voltages from each antenna that have not been subject to any RFI mitigation. An email is then issued to human inspectors 
for a final assessment of the candidate.

\section{Discovery of FRB170827}
\label{Discovery}

Fig. \ref{frbdetection} shows the dynamic spectrum for the FRB at its
detection resolutions of 327.68 $\upmu$s and 97.66 kHz. After correcting for 
the effect of interstellar dispersion, and averaging across the frequency axis, 
the event S/N at this time resolution was 48. Within seconds, it was
evaluated as an FRB candidate by our machine-learning system,
triggering an email alert and a voltage capture of $\sim 270$ ms around the event. 
A failure of the plotting routine (ironically due to the very high S/N of the event, and since corrected) 
meant that FRB170827 was only identified as a \textit{bona fide} FRB three days 
after the event, which limited immediate follow-up at other wavelengths. The result was issued as Astronomer's Telegram ATel 10697 \citep{Farah_FRB170827_atel}.


\begin{table}
  \centering
  \begin{threeparttable}
    \caption{Properties of FRB170827}
    \label{table_properties}
    \begin{tabular}{l c}
      \toprule
      Event UTC & 2017-08-27 16:20:18.1 \\
      Fan-beam number & 92 \\
      S/N (detection fan-beam) & 48 \\
      S/N (coherently dedispersed) & 110 \\ 
      Sampling time & 327.68 $\upmu$s\\
      Detection Filter & 1 (655.36 $\upmu$s width) \\
      
      Best-fitting $\alpha$ (h:m:s) & 00:49:18.66 (J2000)\\
      Best-fitting $\delta$ (d:m:s) & $-$65:33:02.5 (J2000)\\
      Galactic longitude $l$ & 303.29$^{\circ}$ \\
      Galactic latitude $b$ & $-$51.58$^{\circ}$ \\
      $S_{\mathrm{peak}}$ (lower limit)\tnote{\textasteriskcentered} & $60\pm 20$ Jy \\
      Observed Fluence (lower limit)\tnote{\textasteriskcentered} & $20\pm 7$ Jy\,ms\\
      Width (at 10$\%$ maximum) & 400 $\pm$ $10$ $\upmu$s \\
      Refined DM & 176.80 $\pm$ 0.04 pc\,cm$^{-3}$\\
      Galactic DM (NE2001) & 37 pc\,cm$^{-3}$ \\
      Galactic DM (YMW16) & 26 pc\,cm$^{-3}$ \\
      \bottomrule
    \end{tabular}
    \begin{tablenotes}
      \item[\textasteriskcentered] \footnotesize{Corrected for the known position of the FRB within the primary beam pattern in the East-West direction, but uncorrected for the (unknown) FRB position in the north-south direction.}
    \end{tablenotes}
  \end{threeparttable}
\end{table}

The event was seen in 2 fan-beams only (Fig.~\ref{frbdetection}),
consistent with celestial sources and the spacing of our fan-beams
on the sky. We searched all other fan-beams for similar events in time
and dispersion measure, finding none above a S/N of 8. The voltage data were particularly
clean around the time of the event, as expected 
in the very early morning hours (local time 2 AM) on site, as
RFI locally is dominated by mobile handset traffic. 

\begin{figure*}
  \begin{center}
    \includegraphics[width=\textwidth]{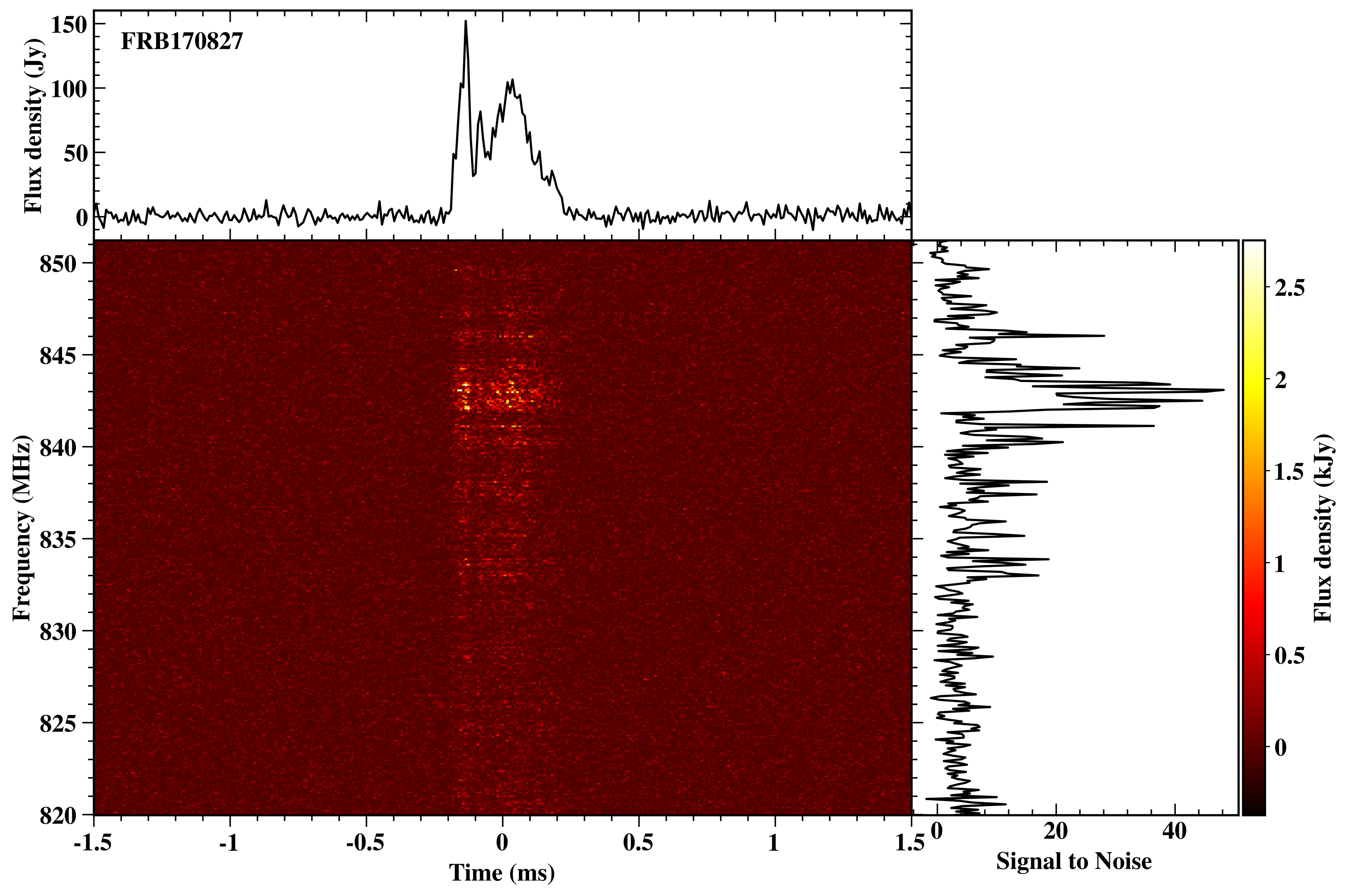}
    \caption{FRB170827 coherently dedispersed with dispersion measure = 176.8 pc cm$^{-3}$.
Structure in FRB170827 is seen at the highest available resolution of the instrument at 10.24 $\upmu$s and 97.66 kHz. The bottom-left panel shows the dynamic spectrum of the event. It shows a region of enhanced emission between 841 and 843 MHz, striations on a scale of 100-200 kHz and ``spiky'' emission features which can be brighter than 1 kJy. The upper panel shows the temporal profile with three major features -- a sharp leading feature, a weak intermediate feature and a broad trailing feature. The right panel shows the time-integrated spectrum of the event as S/N per channel, further illustrating prominent structure on 100-200 kHz scales.}
    \label{FRBhires}
  \end{center}
\end{figure*}

Molonglo’s backend supports a mode in which saved voltages can be read and processed from 
disk, rather than live from the telescope. Voltages allow us to make an improved localisation 
of the FRB, as we can place fan-beams on the sky arbitrarily. We placed 352 fan-beams across 
0.04 deg (beam spacing of $1.1396 \times 10^{-4}$ degrees) covering the two detection fan-beams, compute the S/N of the FRB in each, and fit for the sky position which maximises 
S/N. The best-fitting central 
position for the FRB is $\alpha$ = 00:49:17.68,
$\delta = -$65:33:02.5 (J2000).
Our localisation region is of order 5 arcsec (1-sigma)
in the east-west direction, but is constrained only by the telescope primary 
beam in the north-south direction, with 1-sigma localisation error of 1.2 degrees. 
The localisation arc can be described as:

\begin{equation}
\label{loc_form}
\begin{split}
\delta = -60.71088 -253.7786 \times (\alpha - 0.8) \\
+ 1480.220 \times(\alpha - 0.8)^2,
\end{split}
\end{equation}
where $\alpha$ is in hours, $\delta$ is in deg, and is valid in the $\alpha$ range [0.81, 0.84].
Once the position was optimized, the voltages were streamed through the system 
again to form a tied-array beam on the target position. The signal was 
coherently dedispersed over a range of DMs,  
including a correction for the system gain as a function of frequency (using weights previously obtained from the bright southern pulsar J1644$-$4559), to determine the DM
that maximizes the S/N ratio of the event. We obtain a DM of 176.80 $\pm$ 0.04 
pc\,cm$^{-3}$ and an S/N of 110 (compared to S/N = 48 obtained in the detection fan-beam). 

The event was sufficiently bright in the detection  
fan-beam to saturate $\sim 10\%$ of the 8 bit, high-resolution filterbank data samples 
(resolution 327.68 $\upmu$s, 97.66 kHz) in a $\sim 1$ ms window centered on the event. 
This is the primary reason for the substantial enhancement in the S/N from 48 to 110: the 
much higher dynamic range in the voltage data allows us to fully recover the 
lost flux density. This loss of signal is expected as the parameters used to scale the data stream down to 8-bits are optimised to maximise our sensitivity 
to $\sim 10$ S/N events, while not limiting our ability to discover bright events. 
The S/N improvement is also, to a lesser extent, due to the 
use of coherent dedispersion, localisation of the event within the primary beam, solving 
for an optimal DM and improved time resolution of this narrow event.

Fig.~\ref{FRBhires} shows the event at resolutions of 10.24 $\upmu$s and 97.66 kHz, after 
coherent dedispersion and bandpass correction. 


The DM estimate for the Galaxy contribution along the FRB's line of sight is 37 pc\,cm$^{-3}$ using the 
\textsc{NE2001} model \citep{NE2001} and 26 pc\,cm$^{-3}$ using the \textsc{YMW16} 
model \citep{YMW16}. This results in an average DM 
excess of $\sim 145$ pc\,cm$^{-3}$ for the FRB, the lowest value for any FRB published.  
The upper limit on the DM-inferred redshift is thus $\lesssim
0.12$ \citep{Inoue}. 
The observational properties of FRB170827 are listed in Table~\ref{table_properties}.

\subsection{Possible SMC Origin?}
\label{SMC_origin}

\begin{figure}
  \begin{center}
    \includegraphics[width=\columnwidth]{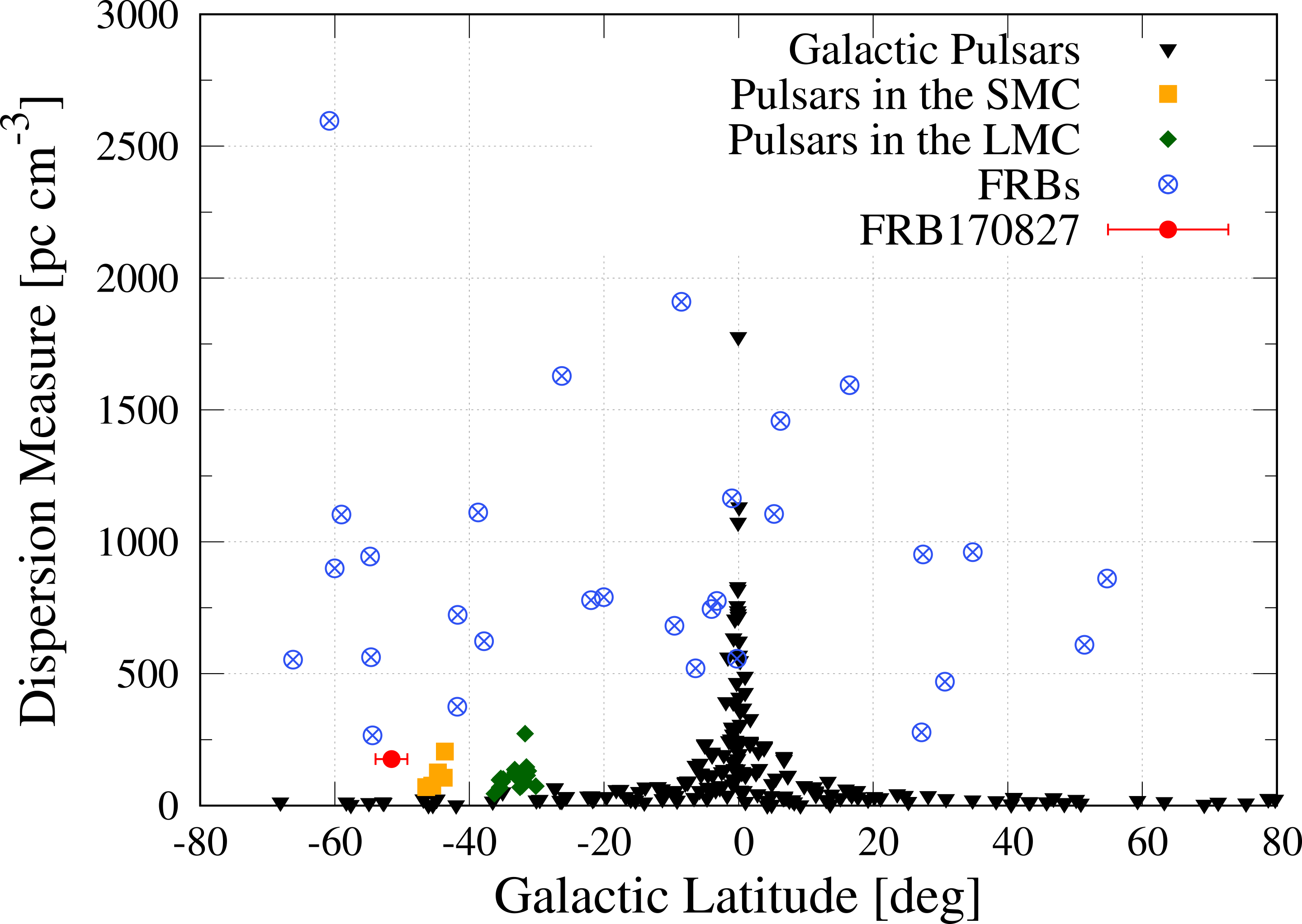}
    \caption{DM versus Galactic latitude plot for published FRBs and
      Milky Way, LMC and SMC pulsars. Galactic pulsars are shown in black, while pulsars in the LMC and SMC are shown in orange and green respectively. FRBs available from FRBcat are displayed in blue. FRB170827 is shown in red and has a DM which is similar to pulsars in the SMC, and a sky position
      $\approx$7 degrees directly north of the SMC, such that the extension of the localisation arc for the FRB source passes through the SMC centre. Nevertheless, we argue in section \ref{SMC_origin} that it is unlikely that the FRB was a giant pulse emitted by a pulsar in the SMC that was detected in a far sidelobe.}
    \label{DMGb}
  \end{center}
\end{figure}

The boresight position of FRB170827 lies $\approx 7$ degrees 
($5\times$ the primary beam half-width half-power) north of the Small 
Magellanic Cloud (SMC), such that the extension 
of the localisation arc (given by Eq. \ref{loc_form}) southward intersects with the central regions of the SMC.
The DM of the event is similar to the DM
of pulsars in the central regions of the SMC, leading us to question whether 
the source could be in the SMC and we have seen a bright event in a
side-lobe. In Fig. \ref{DMGb} we show the DM distribution as a function of Galactic latitude of FRBs
published to date, compared with pulsars in the Milky Way, and in the
Small and Large Magellanic Clouds respectively, showing that FRB170827
overlaps in DM with pulsars in the SMC. Tests of the scenario that FRB170827 
is a far sidelobe detection were carried out in the week 
following the event, in which the telescope was moved 7 degrees south
of the bright southern pulsar Vela, searching for single pulse
events. We found that 7 degrees off boresight to the south of Vela,
occasional pulses could be seen from Vela at a S/N of a few percent of
the boresight S/N. The pulses were detected in one or two adjacent fan-beams, making them otherwise indistinguishable from boresight detections. We estimate that if the source were actually in the
SMC, it could be detected at boresight with a S/N $\ga$ 1000, and would
have a peak flux density of $\ga 2$ kJy. 
It is likely that a source capable of such pulses in the SMC 
would have given off large numbers of fainter pulses that should have been
seen in extant SMC surveys \citep[e.g.][]{Manchester2006}. 
We therefore consider it unlikely that FRB 
170827 resides in the SMC. Pulsar wind nebulae were also searched for along 
the localisation arc, and none were found\footnote{Roberts, M.S.E., 2004, 
`The Pulsar Wind Nebula Catalog (March 2005 version)', McGill University, 
Montreal, Quebec, Canada (available on the World-Wide-Web at 
\url{http://www.physics.mcgill.ca/~pulsar/pwncat.html}).}.

\section{Analysis of FRB170827}
\label{analysis}
Voltage data were captured for 270 ms encompassing the event, and allowed us
to examine the FRB's temporal and frequency structure with much higher
resolution than for all of the non-repeating published FRBs. The dynamic spectrum of the FRB displayed in Fig.~\ref{FRBhires} shows a region of enhanced emission between 841 and 843 MHz, striations on a scale of 100 to 200 kHz and ``spiky'' emission features which can be brighter than 1 kJy. Interestingly, the latter two features are similar to what is seen in FRB150807 \citep{Ravi2016_science}. 

\subsection{Spectral Modulation}
\label{spectral_acf}
Point-like radio sources scintillate due to propagation through inhomogeneous dispersive media. 
To measure the scintillation effects, 
we construct 
the frequency auto-covariance function (ACF) of the spectrum $S(\nu)$:

\begin{equation}
\label{ACF}
A(\Delta\nu) =  \frac{1}{N}\sum_{\nu}\Delta S(\nu)\Delta S(\nu+\Delta \nu),
\end{equation}

\noindent where $\Delta S(\nu) = S(\nu) - \bar{S}$,  with $\bar{S}$ being the mean flux density, and 
$N$ the number of frequency channels.
The zero lag value, associated with self noise, was excised from the auto-covariance function. 
The ACF was then normalised by its maximum and fitted by a Gaussian function of the form:

\begin{equation}
\label{exp_fit}
\xi(\Delta \nu) = \exp\Big[-\text{b}\Delta \nu^2\Big].
\end{equation}

\begin{figure}
  \begin{center}
    \includegraphics[width=\columnwidth]{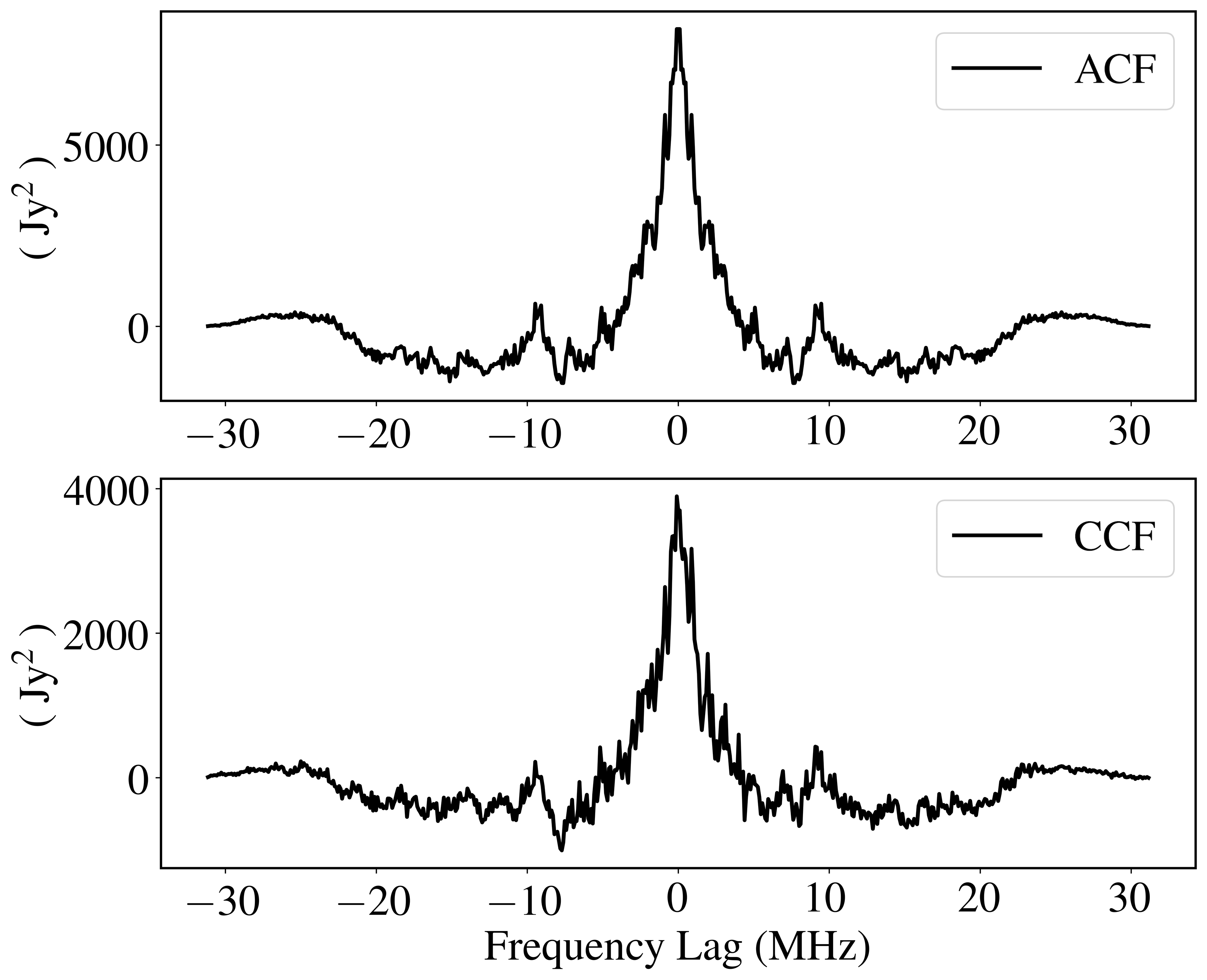}
    \caption{Auto-covariance function (top) of the spectrum of FRB170827, and the 
    cross-covariance function (bottom) of the leading and trailing spectra of the FRB.}
    \label{ACF_CCF}
  \end{center}
\end{figure}

The constructed ACF is shown in Fig.~\ref{ACF_CCF}.
The decorrelation bandwidth, $\Delta\nu_{\textrm{d}}$, is usually defined as the frequency lag where the ACF decays 
to half power \citep{Cordes1986}. The computed 
$\Delta\nu_{\textrm{d}} \sim 1.5$ MHz 
is consistent with 
what is expected along the line of sight, as shown by the \textsc{NE2001} model, where $\Delta\nu_{\textrm{d,NE2001}}\sim$ 0.8 MHz at 835 MHz. 
In Fig. \ref{ACF_CCF}, we show the cross-covariance function (CCF) of the spectra of the leading and trailing features of the temporal profile (see \textsection~\ref{Temporal_structure}). The CCF peaks 
at zero-lag, and shows only modest asymmetry that is consistent with arising from sample variance 
and noise. If the scintillation patterns are identical between the two feature windows, 
the ACF and CCF would have the same shape 
\citep{Cordes1983}. We conclude that the dynamic 
spectrum of FRB170827 is consistent with arising from scintillation.

Another notable feature of the dynamic spectrum of FRB170827 is the 100-200 kHz wide striations seen across the pulse profile. These cannot be 
explained by a passage through the ISM, as they are $\sim$ an order of magnitude narrower than is expected for the line of sight for the NE2001 model. Below, we consider the possibility that they arise in a second scattering screen, well outside the Milky Way.

Interstellar scintillation can amplify pulsars and FRBs
that are otherwise beneath detection thresholds and
make them detectable. FRB170827 was well above
our detection threshold however, and even saturated our
detector system in some channels prior to analysis
of the voltage data. A fainter version of FRB170827 
might have only been visible across $\sim 2$ MHz, and
may have been overlooked by our search algorithms.
Searches for narrow-band FRBs may unveil 
more events if they can still be differentiated from
terrestrial interference.

\subsection{Profile Temporal Structure}
\label{Temporal_structure}
Apart from the repeating FRB, no other FRB has been studied at this timescale due to limitations set 
by DM smearing (e.g. \citealt{Bhandari2018}). Although FRB170827 is much narrower than 
most FRBs \citep{Ravi2017}, this might just be an observational bias 
due to its small level of DM smearing.

\begin{figure}
  \begin{center}
    \includegraphics[width=\linewidth]{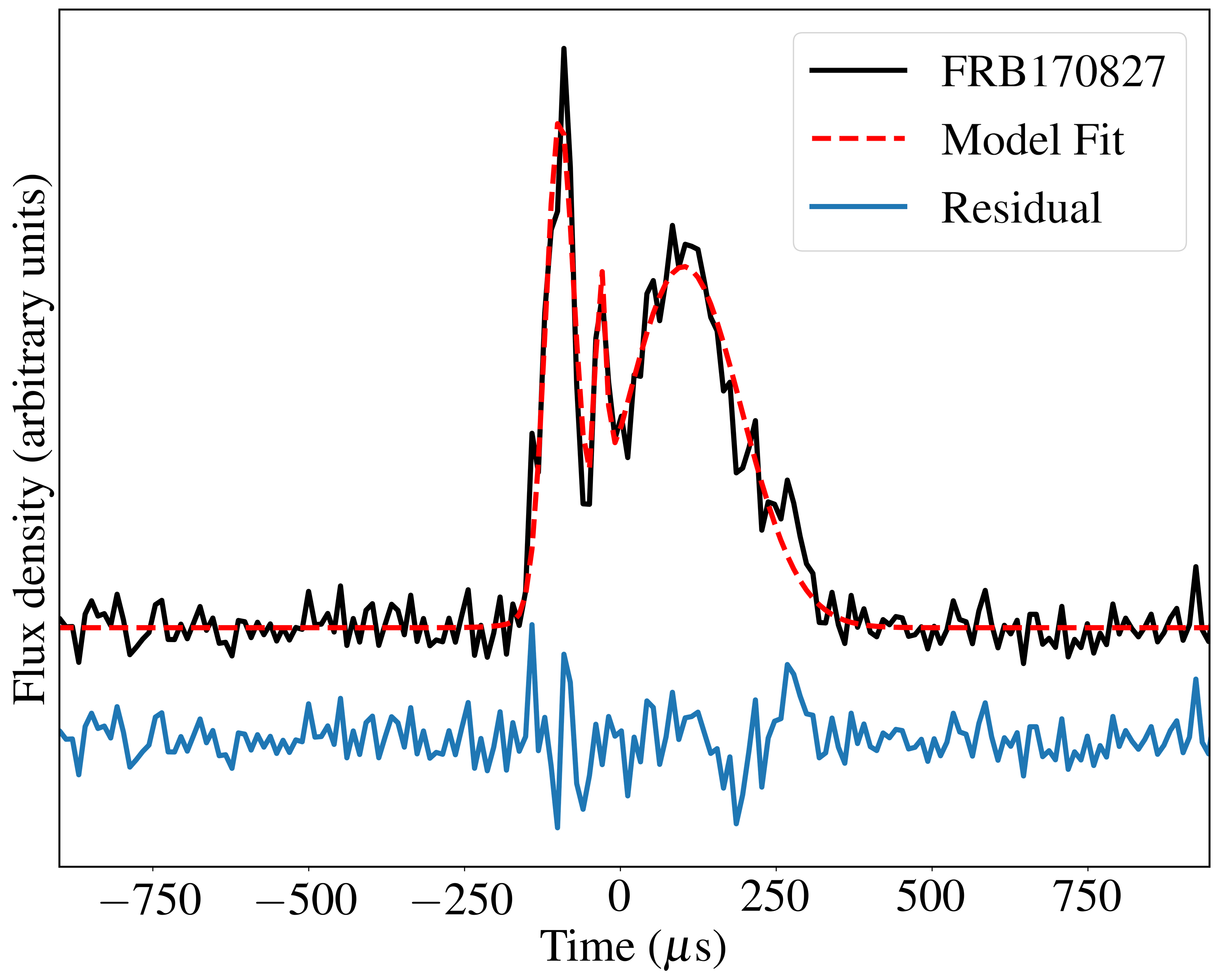}
    \caption{Observed pulse profile (black), model fit (red) and the residual (blue, offset for visibility) for temporal profile of FRB170827. The model consists of 3 Gaussian profiles convolved with a one-sided trailing exponential. The model parameters are listed in Table \ref{model_parameters_table}. The modeling yields a scattering of $4.1\pm2.7$ $\upmu$s.}
    \label{pulse_model}
  \end{center}
\end{figure}

\begin{figure*}
  \begin{center}
    \includegraphics[width=1\textwidth]{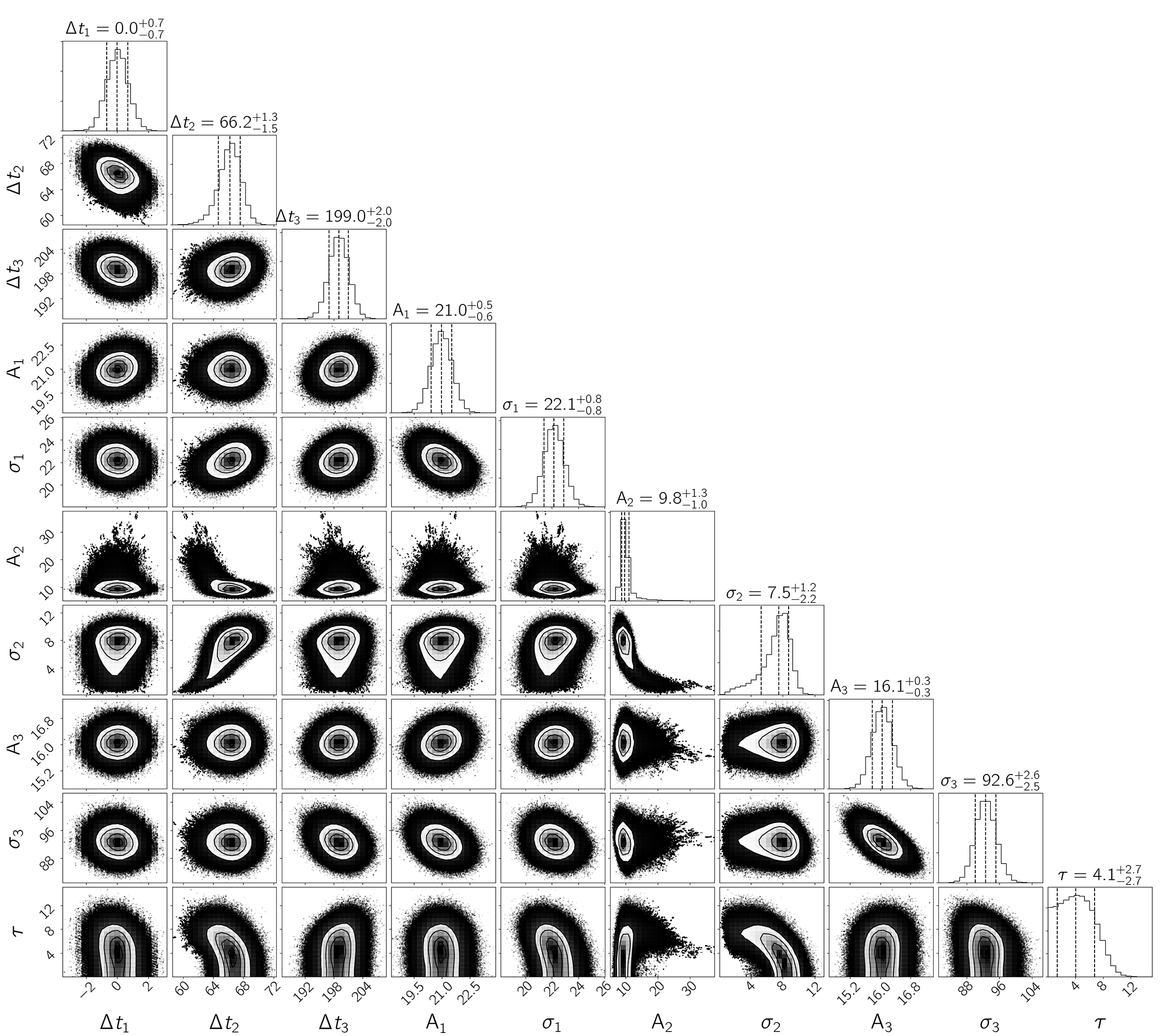}
    \caption{Posterior distributions of the model parameters described in Eq.~\ref{pulse_model}. 
    The dashed lines represent 16, 50 and 84 percentiles in the 1-D histograms.}
    \label{posteriors}
  \end{center}
\end{figure*}

The frequency-averaged pulse profile of the burst shows temporal modulation of the order 
of tens of microseconds and the profile 
can be divided into three different components: a sharp leading feature with a peak flux density 
$\gtrsim 100$ Jy and width $\sim$ 50 $\upmu$s, an intermediate feature, and a trailing feature.
While most FRBs appear to have a single temporal component, some have shown multiple peaks (e.g. 
FRB121002; \citealt{Champion2015}).
%
The pulse profile was fitted 
with a model $\mathbb{G}$ described by the summation of three Gaussian 
profiles, each convolved with a one-sided exponential, of the form:

\begin{equation}
\label{Gaussian-convolved}
\begin{split}
\mathbb{G}_i(t\ |\ \text{A}_i,\Delta t_i,\sigma_i,\tau) = \text{A}_i\times \Bigg[\exp\Big({-\frac{(t-\Delta t_i)^2}{2\sigma_i^2}}\Big)\Bigg] \\
\ast \Bigg[\mathbb{H}(t-\Delta t_i)\exp\Big({-\frac{t-\Delta t_i}{\tau}}\Big)\Bigg],
\end{split}
\end{equation}
\begin{equation}
\textrm{and } \mathbb{G} = \mathbb{G}_1 + \mathbb{G}_2 + \mathbb{G}_3,
\end{equation}

\noindent where $t$ is time in microseconds, $\tau$ is the scattering timescale at 835 MHz, 
$\mathbb{H}(t)$ is a Heaviside unit step function, $\Delta t$ is the time offset of 
a Gaussian relative to the first, $\sigma$ is the Gaussian dispersion width, and the asterisk denotes convolution.
The optimal parameters of the model were obtained using Markov chains Monte Carlo (MCMC)\footnote{Using the 
\textsc{python} package \textsc{emcee} \citep{emcee_Foreman-Mackey}}.
Fig. \ref{pulse_model} shows the best-fitting model $\mathbb{G}$ overlaid on the data, and 
the optimal model parameters are presented in Table~\ref{model_parameters_table}. We show the estimated 
posterior distributions for the model parameters in Fig.~\ref{posteriors}
\footnote{This figure is made using the public
python package \textsc{corner} \citep{corner}.}.

We estimate the scattering timescale $\tau$ to be $4.1 \pm 2.7$ $\upmu$s. This corresponds to frequency structures on a scale of $\sim (2\pi \tau)^{-1}$, consistent with the striations on the 100-200 kHz scale seen in the dynamic spectrum.

In the following sub-section, we assume that this fine structure is induced by a second scattering screen, external to the Milky Way. 
We note that $\tau$ is poorly constrained on the lower end of the posterior distribution, 
and the data are consistent with no scattering at the 2-sigma level. In this case, the structure may be intrinsic to the emission process of the source and in what follows it is important to differentiate between intrinsic emission processes and propagation effects.

We note that if FRB170827 had been recorded with a low-bit digitization system, as was the case, e.g., for the Lorimer burst \citep{Lorimer2007}, the burst presented here would have a much smoother spectrum. Owing to our high dynamic range 8-bit voltage recording, we were able to uncover the wealth of spectral features. If these features are intrinsic to the source, the emission process might be very different to what it is in (for example) the giant pulses from the Crab pulsar \citep[see e.g.,][and references therein]{Hankins2016} which show relatively broad-band features at similar radio frequencies. Detailed comparison to Crab giant pulses is made difficult by lack of polarisation and coarser time resolution for our FRB.


\subsection{Two Screen Model}
\label{2_screen}

The dynamic spectrum of FRB170827 shows scintillation on two frequency scales: 
broadband features explained by a passage through the turbulent ISM, 
and a finer structure with striations 1-2 frequency channels (100-200 kHz) wide, hinting 
at the presence of another scattering screen along the FRB's path from the host to the observer. 
In this section, we assume that both spectral modulation scales arise from scintillation induced 
by two scattering screens, a Galactic screen, and another closer to source.

We model the near screen to be placed $\sim$ 1 kpc from the observer. 
The second screen, in order to give rise to the fine spectral modulation, should be sufficiently  distant 
that it is not resolved by the Galactic screen. Scintillations from a far screen will only be apparent 
if the incident wave field is spatially coherent across the transverse extent of the scattering disk of the near screen. This scale 
is of order the screen's refractive scale, $r_{\textrm{ref}}$, defined as:
\begin{equation}
\label{r_ref}
r_{\textrm{\tiny{ref}}} = \theta_{\textrm{d}} \times D = \frac{r_{\textrm{\tiny{F}}}^2}{s_{\textrm{0}}},
\end{equation}
where $\theta_{\textrm{d}}$ is the angular radius of the source's broadened image, $D$ is the distance to 
the scattering screen, $r_{\textrm{\tiny{F}}}$ is the Fresnel scale, and $s_{\textrm{0}}$ is the field coherence 
scale \citep{handbook}.

We can obtain a first order estimate of the coherence scale of the much further screen by estimating the scale on which 
its average visibility declines. For a source at an angular diameter distance $D_\textrm{S}$ from the observer and an angular 
diameter distance $D_\textrm{LS}$ from the (more distant) lensing screen, this scale is:
\begin{equation}
r_0 = s_{\textrm{0,\tiny{FAR}}} \times \frac{D_\textrm{S}}{D_\textrm{LS}},
\end{equation}
\citep[e.g.][]{Macquart&Koay2013}. The scintillation bandwidth $\Delta \nu_\textrm{d}$ is related to the coherence scale of the scattering screen $s_{\textrm{0}}$ by:
\begin{equation}
\sqrt{ \frac{\Delta \nu_\textrm{d}} {\nu}} =  \frac{s_{\textrm{0}}}{r_{\textrm{\tiny{F}}}},
\end{equation}
\citep{handbook}.
For the Galactic screen, we have:
\begin{equation}
r_\textrm{\tiny{F,ISM}} = \sqrt{\frac{D_\textrm{ISM}}{k}},
\end{equation}
where $k$ is the radiation angular wavenumber. Taking into account the curved geometry of spacetime, the coherence scale of the far screen is:
\begin{equation}
\label{r_F_FAR}
r_\textrm{\tiny{F,FAR}} = \sqrt{\frac{D_\textrm{eff}}{k(1+z_l)}} ,
\end{equation}
where $D_\textrm{eff} = D_\textrm{L} D_\textrm{LS}/D_\textrm{S}$ and $z_l$ are the effective distance and redshift 
of the scattering material respectively \citep{Macquart&Koay2013}. $D_\textrm{L}$ is the angular diameter distance 
from the observer to the scattering screen.

The condition that the more distant screen is not resolved by the Galactic screen is:
\begin{equation}
\label{condition}
r_0 \gtrsim r_{\textrm{ref,\tiny{ISM}}}.
\end{equation}
Using Eqs.~\ref{r_ref} to \ref{r_F_FAR}, Eq.~\ref{condition} reduces to:
\begin{equation}
\frac{\Delta \nu_\textrm{d,FAR}\Delta \nu_\textrm{d,ISM}}{\nu^2}
\gtrsim
\frac{D_\textrm{LS}D_\textrm{ISM}}{D_\textrm{S}D_\textrm{L}} (1+z_l) .
\end{equation}

For our values of $\Delta \nu_\textrm{d,ISM} = 1.5$ MHz, $\Delta \nu_\textrm{d,IGM} = 0.1$ 
MHz, $D_\textrm{ISM} = 1$ kpc, and $D_\textrm{S} = 500 $ Mpc, we get $D_{\textrm{LS}} 
\lesssim 60$ Mpc. The second screen could therefore lie in the Intergalactic Medium, but is 
also consistent with a turbulent screen close to the FRB source in its host galaxy. This 
is similar to the case of FRB110523 \citep{Masui2015}, who found evidence for 
a screen in the host galaxy based on the scattering and polarisation properties 
of the FRB. 

\begin{table}
  \begin{center}
    \caption{Best-fitting parameters to the FRB170827 pulse profile, using 3 Gaussian components with exponential scattering tails (see Eq. \ref{Gaussian-convolved}).
    \label{model_parameters_table}
    $\mathbb{G}_1$, $\mathbb{G}_3$, $\mathbb{G}_3$ represent the leading, 
    intermediate and trailing features of the FRB170827 pulse profile. The
    parameter $\Delta t$ is the time offset of the feature relative to the leading feature, $\sigma$ is the Gaussian dispersion width of each feature. 
    and $\tau$ is the scattering timescale. 
}
    \begin{tabular}{l c c c}
      \toprule
      \ & $\Delta t$ ($\upmu$s) & $\sigma$ ($\upmu$s) & $\tau$ ($\upmu$s)\\
      \toprule
      $\mathbb{G}_1$ & 0.0$^{+ 0.7}_{-0.7}$ & 22.1$^{+ 0.8}_{- 0.8}$ & \multirow{3}{*}{4.1$^{+ 2.7}_{- 2.7}$}\\
      $\mathbb{G}_2$ & 66.2$^{+1.3}_{-1.5}$ & 7.5$^{+ 1.2}_{- 2.0}$ & \\
      $\mathbb{G}_3$ & 199.0$^{+2.0}_{-2.0}$ & 92.6$^{+ 2.6}_{- 2.5}$ & \\
      \bottomrule
    \end{tabular}
  \end{center}
\end{table}

\section{Multiwavelength follow-up}
\label{Multiwavelength}

Follow-up observation of the event were made at the radio wavelengths with UTMOST, ASKAP, and
Parkes, and at optical wavelengths with SkyMapper \citep{SkyMapper}.

\begin{enumerate}

\item UTMOST: FRB170827 occurred while we were running the telescope in FRB
  transit mode after one of the telescope arm drives failed over a
  weekend, so that normal pulsar observing was not possible. 
  As the telescope
  was positioned at $\delta=-65.5$ deg, sources cross through
  the 4 degree primary beam in approximately 40 minutes. 
  This represents 1.5 hours of (serendipitous)
  follow-up of the event in the following 48 hours, and
  before other telescopes could be triggered.
  No candidate bursts were found 60 minutes around the event, 
  nor in the data recorded in the following two days.
  Around 22 hours of follow-up of the 
  FRB were then performed during the period from August 2017 to February 2018, 
  finding no repeat bursts down to a S/N of 9. 

\item Parkes: We searched for repeat bursts using the 20 cm multibeam receiver 
and the BPSR backend, as part of the SUPERB 
project \citep{SUPERB}. The observations were taken in the frequency range 1182-1582 MHz, 
with a usable bandwidth of 320 MHz,
using 1024 channels of width 390 kHz each. Nineteen grid points 
were searched for 300 seconds each along the localisation arc as provided in 
Eq. \ref{loc_form}, spaced
around 20 arcminutes, starting at UTC 2017-08-30 16:39:29. 
No bursts were found with S/N $>$ 8. 

\item ASKAP: 12 hours of follow-up beginning on August 30th, 2017, 22:33:17 UTC, 
centred at the best-fitting position of 
FRB170827's field was performed with a single antenna. The observations were taken at a 
central frequency of 1300 MHz with a bandwidth of 300 MHz. 
No bursts 
were detected above a S/N of 9.5, corresponding to a limiting fluence of 
22 Jy\,ms, at the 1.26 ms time resolution.

\item SkyMapper: Several nights of imaging data were taken in the week 
after the event (2017-08-29, 2017-08-30, 2017-09-01). 
Images were taken in the $uvgriz$ bands (100 s exposures) 
with photometric depth limits of $u = 18.1$, $v = 18.5$, $g = 20.5$, 
$r = 20.3$, $i = 19.5$ and $z = 18.7$, at the 95\% upper limit provided 
by the SkyMapper Transient Survey Pipeline \citep{Scalzo2017}. 
The follow-up fields were centred around the FRB 
coordinates, extending north-south to cover the 2-sigma error regions 
(i.e. 4.8 degrees). We carried out two different follow-up modes over two nights: 
the first consists on images centred on the FRB position with multiple visits 
with slight pointing offsets, and the second takes images of the SkyMapper field covering 
the FRB localisation and the 2-sigma regions.

Eight galaxies were found in the 6dFGS catalog \citep{6dFGS} lying along the localisation arc of the FRB, 
in the redshift range $0.005 < z < 0.087$. They lie in the magnitude ranges 
$13.6 < r_F < 16.3; 12.8 < B_J < 16.3$, and are mostly disk galaxies. 
Reference images were taken on 2015-07-06 ($\sim$2 years ago) and on 
2017-09-02 ($\sim$4 days after the first epoch of the FRB followup observations). 
Only five galaxies are well placed on the CCD images to permit proper
processing with our image subtraction procedure using both the newly acquired
data and the reference images. No transient or variable candidates were
detected in the SkyMapper data along the localisation arc.

\end{enumerate}

\section{Discussion and Conclusions}
\label{discussion}

In this paper, we reported a Fast Radio Burst (FRB170827) discovered in near real-time at the Molonglo radio telescope. 
This demonstrated our ability to trigger voltage capture for a new FRB with a low-latency 
machine-learning-based discovery system. The discovery allowed, and for the 
first time, to perform coherent dedispersion of a burst after its blind detection, unveiling 
temporal structure that would not have been otherwise observed.

The full width of the burst is 400 $\upmu$s, and it has a DM of 176.80 $\pm$ 0.04 pc\,cm$^{-3}$ (after coherent dedispersion), which is 
the lowest known DM of the FRB population. The Milky Way contribution along 
the line of sight is $\sim$ 37 pc\,cm$^{-3}$ (\textsc{NE2001}) or $\sim$ 26 pc\,cm$^{-3}$ (\textsc{YMW16}), leaving an excess of $\sim$ 140 or 150  
pc\,cm$^{-3}$, and limiting its redshift to $z < 0.12$, potentially placing it closer 
than the repeating FRB (FRB121102, for which the host galaxy is at $z = 0.193$). It 
has an observed fluence of $>$ 20 $\pm$ 7 Jy\,ms, placing it amongst the brightest FRBs 
found to date. 

The dynamic spectrum of the FRB shows spiky emission features of up to 1 kJy at 10.24 $\upmu$s and 97.66 kHz resolution, similar to the very bright (50 Jy\,ms) FRB150807 \citep{Ravi2016_science}, who reported bright spikes of over 1 kJy at a resolution of 64 $\upmu$s and 390 kHz.

The temporal profile of the burst shows three components, the narrowest of 
which is $\sim$ 30 $\upmu$s. FRB170827 shows spectral modulations on two frequency scales of $\sim$ 1.5 MHz and 0.1 MHz. 
The latter is based on the scattering of the event of $4.1 \pm 2.7$ $\upmu$s,
obtained via maximum likelihood fitting of the three burst components.   
These two scales are clearly visible in the dynamic spectrum of the event which is dominated by a bright region of emission between 841 and 843 MHz, but with weaker, patchy emission across the entire band. The present patchy emission
is similar to what is seen in the repeating FRB (FRB121102; \citealt{Michilli2018}) at 4.5 GHz, FRB110523 \citep{Masui2015} at 800 MHz 
and in FRBs found at ASKAP (Shannon et al. in prep) at 1.4 GHz. A decorrelation bandwidth of $\sim$0.8 MHz (at 835 MHz) is expected due to the ISM at the position of FRB170827 ($(l,b) = (303.29^\circ, -51.58^{\circ})$) using the \textsc{NE2001} model, which is consistent with the larger of these scales. The 0.1 MHz scale striation is significantly lower than can be accounted for 
from the ISM, and we speculate it could arise in the host galaxy of the FRB, similarly to the high-RM FRB110523 and the repeating FRB (FRB121102). 

Microstructure visible in the temporal profile of FRB170827 is very similar to that seen in the repeating FRB121102. This and other similarities to the repeater has strongly motivated a follow-up campaign for repeat bursts, currently being conducted at UTMOST at 835 MHz to a fluence limit of approximately 5 Jy\,ms and Parkes at 1.4 GHz to a limit of approximately 0.5 Jy\,ms. 


We are currently upgrading the interferometer's second arm (UTMOST-2D project), and our FRB detection and voltage capture system will enable us to perform FRB host galaxy localisation to a few arcsec accuracy from single FRB events by recording voltages from all the array elements, such as was achieved for FRB170827.

\vskip 1.0 truecm
\section*{Acknowledgments}
The authors would like to thank the anonymous referee for useful comments and suggestions. 
The Molonglo Observatory is owned and operated by the University of Sydney, with support from the 
School of Physics and the University. The UTMOST project is also 
supported by the Swinburne University of Technology.
We acknowledge the Australian Research Council grants CE110001020 (CAASTRO) and the Laureate Fellowship 
FL150100148. We thank Dave Temby, Glen Torr, Glenn Urquhart, Simon Jordan for cheerfully keeping the 
UTMOST facility performing so smoothly. 
We thank Kiyoshi Masui for helpful discussions. 
The Parkes radio telescope is part of the Australia Telescope National Facility, which is funded by the 
Commonwealth of Australia for operation as a
National Facility managed by CSIRO. 
The ASKAP is part of the Australia Telescope National Facility, which is managed by CSIRO. Operation of ASKAP is funded by the Australian Government with support from the National Collaborative Research Infrastructure Strategy. ASKAP uses the resources of the Pawsey Supercomputing Centre. Establishment of ASKAP, the Murchison Radio-astronomy Observatory and the Pawsey Supercomputing Centre are initiatives of the Australian Government, with support from the Government of Western Australia and the Science and Industry Endowment Fund. We acknowledge the Wajarri Yamatji people as the traditional owners of the Observatory site.
ATD is supported by an ARC Future Fellowship grant FT150100415. 
Parts of this work was performed on the gSTAR national facility at Swinburne
University of Technology. 
SkyMapper is owned and operated by The Australian National University's Research School of Astronomy and Astrophysics. The national facility capability for SkyMapper has been funded through ARC LIEF grant LE130100104 from the Australian Research Council, awarded to the University of Sydney, the Australian National University, Swinburne University of Technology, the University of Queensland, the University of Western Australia, the University of Melbourne, Curtin University of Technology, Monash University and the Australian Astronomical Observatory. 

\bibliographystyle{mnras}
\bibliography{bibliography}

\bsp	
\label{lastpage}
\end{document}